# Real-time full field measurements reveal transient dissipative soliton dynamics in a mode-locked laser

P. Ryczkowski[1†], M. Närhi[1†], C. Billet[2†], J.-M. Merolla[2], G. Genty[1], J. M. Dudley[2]*

1   Laboratory of Photonics, Tampere University of Technology, Tampere, Finland.

2   Institut FEMTO-ST, UMR 6174 CNRS-Université Bourgogne Franche-Comté, Besançon, France.

† Equal contributions

* Corresponding author john.dudley@univ-fcomte.fr

**Dissipative solitons are remarkable localized states of a physical system that arise from the dynamical balance between nonlinearity, dispersion and environmental energy exchange. They are the most universal form of soliton that can exist in nature, and are seen in far-from-equilibrium systems in many fields including chemistry, biology, and physics. There has been particular interest in studying their properties in mode-locked lasers producing ultrashort light pulses, but experiments have been limited by the lack of convenient measurement techniques able to track the soliton evolution in real-time. Here, we use dispersive Fourier transform and time lens measurements to simultaneously measure real-time spectral and temporal evolution of dissipative solitons in a fiber laser as the turn-on dynamics pass through a transient unstable regime with complex break-up and collision dynamics before stabilizing to a regular mode-locked pulse train. Our measurements enable reconstruction of the soliton amplitude and phase and calculation of the corresponding complex-valued eigenvalue spectrum to provide further physical insight. These findings are significant in showing how real-time measurements can provide new perspectives into the ultrafast transient dynamics of complex systems.**

**INTRODUCTION**

When operating in their steady-state regime, mode-locked lasers produce highly regular stable pulse trains which have found widespread application in many fields of science (*1-3*). However, the mode-locked laser is also well-known to exhibit a rich variety of *unstable* dynamical phenomena when detuned from its steady-state, or as stable mode-locking builds up from noise (*4-6*). These transient instabilities are a subject of particular interest in passively mode-locked fiber lasers, since the interplay of the fiber nonlinearity and dispersion with cavity gain and loss yields a rich landscape of *dissipative soliton* dynamics (*7-10*). Dissipative solitons are in fact the most universal class of localized soliton state in physics, existing for an extended period of time in the presence of nonlinearity and dispersion, even while parts of the structure can experience gain and loss. In contrast to solitons in energy-conserving systems, dissipative solitons exist in systems far from equilibrium, and are dynamical objects that display highly non-trivial behaviour (*7*).

Although there is an extensive theoretical literature on dissipative solitons in fiber lasers, experimental studies have been more limited, and often restricted to measurements using only fast photodetectors (*6,9,11-13*). These experiments have certainly provided insight into unstable modes of laser operation, yet the limited temporal resolution of the detectors used has generally prevented any detailed study of the underlying soliton dynamics.

Recent years, however, have seen dramatic advances in techniques capable of real-time measurements of non-repetitive optical signals (*14-25*). The first method to see widespread use was the dispersive Fourier transform (DFT) technique for real-time spectral measurements at the MHz repetition rate typical of mode-locked lasers (*14*). The DFT was first used to study fiber instabilities such as rogue wave dynamics, modulation instability and supercontinuum generation (*15-18*), and more recently has been applied to study the spectral properties of mode-locked lasers. The ability to resolve the spectrum after each cavity round trip in a laser has revealed novel decoherence and soliton explosion dynamics in a fiber laser (*19,20*), and wavelength evolution dynamics in a Kerr-lens Ti:Sapphire laser (*21*). In fact, soliton dynamics in the Kerr-lens

Ti:Sapphire laser have also been studied using a modified version of the DFT adapted for real-time spectral interferometry (*22*). In parallel with these advances in real-time spectral measurements, the development of a temporal analog of a spatial thin lens has resulted in real-time pulse intensity measurements in with sub-picosecond resolution (*23*). The time-lens has been used in studies of incoherent soliton propagation in optical turbulence (*24*), and stochastic breather emergence in modulation instability (*25*).

In this work, we combine, for the first time, real-time spectral DFT and time lens techniques to perform simultaneous measurements of the spectral and temporal profiles of dissipative solitons evolving during the turn-on phase of a mode-locked fiber laser. Our measurements are made with sub-nm and sub-ps resolution, providing a unified snapshot of how the solitons evolve from round trip to round trip, and directly revealing a range of highly complex dynamics typical of dissipative solitons. In addition, phase retrieval allows the reconstruction of the full field (intensity and phase) of the measured pulses, allowing us to determine the corresponding complex eigenvalue spectrum of the pulses in different dynamical regimes. These results are significant in providing a unique picture of the internal evolution of fiber laser dissipative solitons, and we anticipate their application in the optimization and design of lasers with improved stability characteristics. More generally, we believe that our results will stimulate the widespread use of simultaneous temporal and spectral characterization as a standard technique for the study of ultrafast complex optical systems.

**RESULTS**

Our experiments studied the transient turn-on dynamics of a saturable absorber mode-locked fiber laser configured (in stable operation) to generate pulses of ~4.5 ps duration at 1545 nm with 20 MHz repetition rate (i.e. cavity round trip time of 50 ns). The laser cavity has net anomalous dispersion such that in steady-state operation the circulating pulses are close to fundamental solitons, and indeed frequency-resolved optical gating measurements confirmed that the stable output pulses were well-fitted by a hyperbolic secant profile and were transform limited with a uniform temporal phase (*26*).

We characterized the turn-on dynamics using both direct photodiode detection as well as simultaneous DFT and time-lens measurements (see Materials and Methods for full details). Fig. 1 shows results using direct photodiode detection. Specifically, Fig. 1(a) plots the envelope of the pulses measured with reduced detection bandwidth to capture a time-window of 130 ms, corresponding to 2.6 million cavity round trips. The figure clearly reveals a ~100 ms transient regime of Q-switched mode-locked operation before the appearance of a stable pulse train. Figure 1(b) shows a separate measurement of a portion of the transient regime as indicated, but here using 30 GHz detection bandwidth so that it is possible to resolve a series of ~10 µs bursts separated by a ~70 µs period. The expanded view in Fig. 1(b) shows how under each burst, an irregular train of mode-locked pulses (at 50 ns period) is present. In contrast with this unstable regime, Fig. 1(c) shows measurements in the stable mode-locked regime using 30 GHz detection, to illustrate a regular pulse train of constant intensity. Of course, such complex transient dynamics have been seen in both numerical and experimental studies of a range of other passively-mode-locked lasers (*27-34*), but we include these measurements here for completeness. The key point is that, even with a detection bandwidth of 30 GHz, the temporal width of the pulses seen in the photodiode intensity record is ~30 ps, precluding any detailed study of the underlying soliton dynamics in the transient regime. It is this limitation that we overcome with our simultaneous real-time DFT and time-lens measurements.

Figure 2 shows the setup used. Light from the fiber laser output was split into two paths and sent to the real-time DFT and time-lens acquisition arms. Real time spectral measurements used a standard DFT set up (*18*) with spectral resolution of ~0.3 nm. The time lens set-up was similar to that described in Ref. (*25*) and was capable of real-time measurements of intensity profiles up to 60 ps duration with an effective temporal resolution of 400 fs. Both DFT and time lens signals were recorded using a high storage capacity digital oscilloscope, with data acquisition triggered by the time-lens signal detected after the laser was turned on. We could simultaneously measure spectral and temporal intensity profiles over a maximum of 400 cavity roundtrips (see Materials and Methods).

We recorded multiple data sequences to examine the spectral and temporal soliton dynamics both in the Q-switched mode-locked and the stable-modelocking regime. We begin in Fig. 3(a) by showing results for the stable modelocking regime where as expected the spectral (i) and temporal (ii) intensity profiles are constant with roundtrip. The extracted spectral and temporal widths and the integrated pulse energy are shown as the right subfigures; the small (noise) variations in these figures is of the order of the experimental resolution of the spectral and temporal measurements.

Although the ability to measure both temporal and spectral intensity in real time is itself a highly significant advance, we extend this technique further using Gerchberg-Saxton phase-retrieval (*35,36*) to recover the corresponding complex electric field of each pulse (see Materials and Methods). These results are shown in Fig. 3(a-iii) where we plot the retrieved intensity (left axis) and phase (right axis) revealing the soliton characteristics of near-uniform phase. Access to the full complex electric field allows us to calculate the associated wavelength-time spectrogram, and these results are shown in Fig. 3(a-iv), confirming the localized soliton-like nature of the pulses. Fig 4 shows similar results, but from a data sequence just prior to the regime of stable modelocking. Here, over a relatively small number of round trips, we see significant modulation in both the measured spectral and temporal amplitudes, as well as energy variation of ~30%. This "breathing" of the intracavity pulses just prior to the onset of stability is a known property of dissipative soliton dynamics in mode-locked lasers (*21*), but our results are the first to be able to experimentally characterize it completely in both the time and frequency domains.

Dissipative solitons can display a much richer range of interaction dynamics (*7-9*), and Fig. 4 shows examples of this behavior in the evolution of pulses under a transient Q-switched burst. To capture different bursts at different times we used a variable hold off time for our measurement window to scan various points of the transient regime. In fact, we found that whilst the general dynamics were similar for all burst durations (i.e. showing the emergence of multiple pulses from an initial noisy field followed by subsequent decay) the detailed evolution dynamics varied significantly for different bursts. Indeed, our measurement set-up allowed us to reveal previously unobserved regimes of dissipative soliton propagation and interaction under the transient envelope.

Remarking firstly that the spectral and temporal evolution with round trip is plotted from top to bottom, two typical results are shown in Fig. 4(a) and 4(b) to illustrate the different types of behavior observed. Also note that energy is not constant in the transient turn-on regime (*12*). We first discuss Fig. 4(a) which shows an initial noisy field splitting into 3 distinct pulses (of duration ~5 ps) that propagate coherently together over ~100 roundtrips. Each pulse displays a well-localized temporal intensity peak, and the three pulses are mutually coherent as is seen by the fact that there is distinct modulation in the corresponding spectrum. Significantly, the pulse separation does not vary over ~100 roundtrips as the pulses evolve without seeming to interact before decaying. Figure 4(b) shows a qualitatively different case. Here, two pulses emerge, but rather than propagating without interacting, they undergo attraction and eventually collide, displaying a more complex nonlinear phase profile at this point.

Although a detailed study of these dynamics is beyond the scope of this paper, we remark that analysis of these results (and a more extensive data set of over 100 similar cases) reveal that whether or not the emergent solitons evolve without interacting or collide depends on whether or not the pulses have the same central frequency. The central frequencies of the pulses can be clearly seen on the spectrogram plots, and in Fig. 4(a) we see how the three (non-interacting) pulses have slightly different central frequencies whereas in Fig. 4(b) the two pulses that collide have the same central frequency and similar amplitudes. These results are consistent with previous studies of bound soliton dynamics (*37-39*). Figure 4(c) on the other hand shows another example of burst evolution where a combination of the above dynamics are observed with the emergence of 3 solitons. Here, the two solitons with the same central frequency attract each other and collide after ~100 roundtrips, whereas the remaining soliton with lower amplitude and a slightly different frequency propagates without interaction before eventually decaying (*40,41*).

The ability to completely reconstruct the complex electric field from the measured pulse intensity and phase is highly significant, and indeed allows calculation of derived quantities that yield additional and important physical insights. In particular, it is possible to apply numerical techniques from scattering theory (*42-45*) to calculate a nonlinear Fourier transform that yields spectral eigenvalue portraits of the pulse structures which can then be compared with the known signatures of ideal solitons (*42*). Although the use of such scattering theory in fiber optics has generally been from a theoretical perspective to obtain analytic solutions to the governing partial differential equations, there has also been recent interest in numerical approaches to yield insight into rogue wave dynamics (*46*) and as a novel approach to overcoming channel limits in optical communications (*47*). Significantly, although the theoretical analysis is only strictly valid in an integrable system, numerical computation of the eigenvalue spectrum for the dissipative soliton system considered here (see Materials and Methods) nonetheless yields results that have a clear physical interpretation. This is because we consider evolution regions where the pulse properties are primarily determined by nonlinear and dispersive effects (i.e. we can neglect resonant gain shaping) so the system may be considered to be only weakly non-integrable. In this case the evolution of pulses through the different discrete elements of the cavity can be approximated by an equivalent nonlinear Schrödinger equation (NLSE) with uniform distributed nonlinearity and dispersion, and the nonlinear spectrum can be considered as yielding local eigenvalues that can reveal approximate soliton content (*48*).

The results of this analysis are shown in Fig. 5 where the red dots in each subfigure show the calculated discrete eigenspectrum for the normalized pulse intensity profiles shown as insets. In this case, we associate the stable pulse regime show in Fig. 3(a) with solitons having with eigenvalues at $\xi = \pm 0.5i$ in the complex plane. Based on the measurements in Fig. 3(a), we can then normalize the intensities of measured pulses in the transient regime with respect to these ideal solitons and perform direct scattering analysis to determine the corresponding more complex eigenvalue spectrum (see Materials and Methods).

We first consider the results in Fig. 5 (a) and (b) which correspond to single pulses at two points of the breather cycle shown in Fig. 3(b-iii). In each case the scattering analysis yields one distinct eigenvalue, and we see that as the pulse intensity varies below and above the unity value of a normalized fundamental soliton (dashed black line), the eigenvalue Im($\xi$) also varies below and above the ideal soliton value of ±0.5$i$. In other words, the variation in the eigenvalue reflects the breathing of the pulse properties in this regime.

Figure 5(c) and (d) consider more complex evolution. Figure 5(c) plots the eigenvalue spectrum for the double-soliton pulse shown in Fig. 4(b-iii) where the direct scattering procedure yields two discrete eigenvalues, both with Im($\xi$) ~ ±0.5$i$. Figure 5 (d) shows results for the three-soliton case in Fig. 4(a) and here we see two discrete eigenvalues with Im($\xi$)~±0.5$i$ and a third eigenvalue at a slightly lower value. In all cases, the clear separation of the eigenvalues from the real axis indicates that this particular dynamical regime within the laser bursts prior to steady-state modelocking can be considered as creating discrete pulse complexes, similar to those seen with multiple bound solitons (*7-9*). This is a very important physical insight that the nonlinear Fourier transform reveals directly (*48*).

**DISCUSSION**

Dissipative nonlinear systems can be highly complex, but it is clear that recent years have seen tremendous advances in being able to understand their behavior through novel theoretical approaches and numerical modelling. Of course, a full understanding of such complex dynamical phenomena requires that theoretical and numerical results are carefully compared with experiment, and to this end the measurement technique and phase retrieval analysis reported in this paper make a significant contribution in enabling complete characterization of picosecond dissipative soliton systems.

The results obtained show a wide range of ultrafast temporal and spectral dynamics never before seen directly, and we anticipate that our results will stimulate many future theoretical studies and numerical modelling that can now be compared directly with experiment in a way that was not previously possible.

A further area of particular significance concerns our use of the nonlinear Fourier transform to calculate a local eigenspectrum of the dissipative solitons in the laser. The calculated spectra clearly show the presence of local soliton content in the complex pulse profiles measured, which provides a new window into the physics of the underlying laser dynamics. We also believe our results will motivate interest in much broader applications of the nonlinear Fourier transform in all dissipative systems where field evolution through discrete elements can be modelled approximately by uniformly distributed nonlinearity and dispersion (*49,50*). In optics, we anticipate particular interest in real-time spectral and temporal studies of nonlinear single-pass propagation dynamics in optical fibers such as rogue wave and modulation instability that also display complex transient noise spikes.

**SUPPLEMENTARY MATERIAL**

**MATERIALS AND METHODS**

**Experimental Setup**

The fiber laser used in our experiments was a commercial Pritel FFL-500 model using a linear Fabry-Perot cavity configuration, similar to that described in (*51*) but with a 978 nm pump laser and Er:doped fiber gain medium. Modelocking is sustained by a 2 μm thick bulk saturable absorber (InGaAs on InP substrate) contact bonded to one of the cavity mirrors. The steady-state modelocking dynamics are dominated by soliton propagation effects because of the net anomalous dispersion in the cavity, with the generation of hyperbolic-secant like pulses of flat phase in stable operation.

We implemented the DFT technique using 850 m of dispersion compensating fiber (DCF) with group velocity dispersion coefficient of 100 ps/nm.km and dispersion slope 0.33 ps/nm$^2$.km at 1550 nm. We attenuated the input to the DCF in order to ensure linear propagation, and confirmed the fidelity of the time-stretching technique when the fiber laser was operated in the stable mode-locked regime by comparing the DFT spectrum with that measured using an optical spectrum analyser (Anritsu MS9710B). The real-time DFT signal was measured by a 35 GHz photodiode (New Focus 1474 A) connected to a 30 GHz channel of a real-time oscilloscope (LeCroy 845 Zi-A, 80 GS/s), resulting in a spectral resolution of 0.3 nm.

The time lens measurements used a commercial Picoluz UTM-1500 system described previously in (*24*) with a temporal magnification factor of 76.4. Total accumulated dispersion for the input and output propagation steps was: $D_1$ = 4.16 ps/nm and $D_2$ = 318 ps/nm respectively, with magnification $|M| = D_2/D_1$. The temporal quadratic phase (to reproduce the effect of a thin lens) was imposed through four wave mixing from a pump pulse (100 MHz Menlo C-Fiber Sync and P100-EDFA) with linear chirp accumulated from propagation in a pre-chirping fiber $D_P$. The imaging condition for magnification is $2/D_P = 1/D_1 + 1/D_2$ so that the dispersion for the pump is around twice that of the signal input step. The signal at the time lens output is recorded by a 13 GHz photodiode (Miteq 135GE) connected to the 30 GHz channel of the real-time oscilloscope at a sampling rate of 80 Gs/s, resulting in an effective 400 fs temporal resolution over a (demagnified) time window of 60 ps. The temporal window is determined by the pump pulse duration used in the time lens to impose the required quadratic chirp via four-wave mixing in a highly nonlinear Si waveguide (*52*). In stable mode-locked operation it is possible to synchronize the 20 MHz laser under study with the 100 MHz pump laser, but this is not the case when studying the transient dynamics as there is significant amplitude and phase noise that precludes the detection of a well-defined 20 MHz harmonic for repetition-rate locking. The time lens is therefore operated in asynchronous mode with free-running acquisition triggered by the arrival of the time lens signal, although this limits the number of roundtrips that can be simultaneously measured to ~400 (as there is a walk-off between the Q-switched mode-locked pulses relative to the time lens gate). To obtain representative data sets at different points in the Q-switched mode-locked burst (which typically develops over 200-300 round trips) we performed multiple measurements with different delays between the switch-on time of the fiber laser and the time lens trigger. In this way, we were able to characterize different phases of the pulse evolution in the transient regime. Finally, we note that because the optical path lengths of the DFT and time lens steps were different, the delay between time-lens and DFT records was calibrated (in a separate measurement) by comparing the arrival times on the oscilloscope of a characteristic intensity pattern imprinted onto a CW laser. This allowed us to match without ambiguity the real-time temporal and spectral intensity profiles of the dissipative solitons during their transient evolution phase.

**Phase Retrieval**

Phase retrieval was performed with the Gerchberg-Saxton algorithm (*35*). Aside from a few trivial ambiguities, this algorithm is known to be accurate for pulse retrieval when intensity envelopes are measured both in the spectral and time domains. The algorithm constructs an initial guess for the profile of the electric field using the measured temporal intensity $I_M(t)$ and an initial random phase $\phi_{\text{rand}}(t)$: $E_g(t) = \sqrt{I_M(t)}\, e^{i\phi_{\text{rand}}(t)}$ which is then Fourier transformed to the spectral domain to yield a corresponding initial guess for the spectral amplitude and phase FT $[E_g(t)]$ = $\sqrt{S(\omega)}e^{i\varphi(\omega)}$. The next step involves retaining the calculated spectral phase but replacing the calculated spectral amplitude with the measured spectral amplitude i.e. $\sqrt{S_M(t)}e^{i\varphi(\omega)}$. This updated spectral profile is then transformed back into the time domain where we retain the calculated temporal phase but again replace the calculated amplitude with that from experiment $\sqrt{I_M(t)}$. This procedure is iteratively repeated until the root mean square error between the measurements and retrieved intensity profiles becomes smaller than a chosen value (3 ×10$^{-5}$ in our case). Convergence was improved by applying the measured temporal and spectral constraints only where the measured intensities were well above the noise floor (-20 dB from the maximum). Elsewhere the amplitudes were multiplied with a small constant value of 0.001 forcing them below the noise. If the algorithm was detected to stagnate (no change in the retrieval error before reaching the desired value), a small additional random phase contribution was added. The reliability of the algorithm was tested with simulated pulses with complex properties similar to those seen in experiments, and by performing multiple retrievals on the same experimental data, which all converged to the same results (within the retrieval error).

**Nonlinear Fourier Transform**

The nonlinear Fourier transform (also known as the direct scattering transform) is a mathematical procedure that identifies and quantifies soliton content in a given pulse structure. Significantly, whilst its use in theoretical analysis to obtain closed-form analytic soliton solutions is only strictly valid in an integrable system, computation of the eigenvalue spectrum can be performed numerically for any arbitrary optical field. Of course, the question in this case is how such results should be interpreted, but for a laser where the pulse properties are primarily determined by nonlinear and dispersive effects, the nonlinear spectrum can be considered as yielding local eigenvalues that can reveal approximate soliton content (*48*). This is because if we can neglect strong pulse filtering effects (e.g. from the resonant transition or saturable absorber) then the evolution through the different discrete elements of the cavity can be considered as yielding an average pulse that is a solution to an equivalent NLSE with uniform distributed nonlinearity and dispersion. In fact this approach is the basis of the well-known use of the complex Ginzberg Landau (or Haus Master Equation) model that has proven highly successful in modelling mode-locked lasers (*8,9,53*).

Under these conditions considering a system described by the NLSE in normalised form:

$$i\frac{\partial \psi}{\partial \xi} + \frac{1}{2}\frac{\partial^2 \psi}{\partial \tau^2} + |\psi|^2 \psi = 0 \qquad (1)$$

The associated scattering problem yields the following system (*42*):

$$\frac{\partial v_1}{\partial \tau} + \psi v_2 = \zeta v_1$$
$$\frac{\partial v_2}{\partial \tau} + \psi^* v_1 = -\zeta v_2 \qquad (2)$$

where $v_1$ and $v_2$ are the amplitudes of the waves scattered by the potential $\psi$, and $\zeta$ is the corresponding complex eigenvalue. For our results, stable modelocked operation was assumed to correspond to (intracavity) solitons with $|\psi| = 1$, and the field profiles corresponding to the other pulses analysed were normalised relative to this value. Standard numerical techniques (matrix methods) were used to determine the eigenvalue spectrum (*48*).

**ACKNOWLEDGEMENTS**

**Funding**

This work was supported by the Agence Nationale de la Recherche project LABEX ACTION ANR11-LABX-0001-01, the Region of Franche-Comté Project CORPS and the Academy of Finland (Grants 267576 and 298463).

**Contributions**

All authors participated in the experimental work and data analysis reported, and to the writing and review of the final manuscript. G.G. and J.M.D. planned the research project and provided overall supervision. The authors also thank K. V. Reddy for providing technical details concerning the soliton operating regime of the Pritel laser used in these experiments.

**Competing financial interests**

The authors declare no competing financial interests.

**Corresponding authors**

Correspondence to: John M. Dudley

**REFERENCES**

1. P. M. W. French, The generation of ultrashort laser pulses, *Rep. Prog. Phys.* **58,** 169-267 (1995).

2. U. Keller, Recent developments in compact ultrafast lasers, *Nature* **424**, 831-838 (2003).

3. T. W. Hänsch, Nobel Lecture: Passion for precision, *Rev. Mod. Phys.* **78**, 1297-1309 (2006).

4. D. Abraham, R. Nagar, V. Mikhelashvili, G. Eisenstein, Transient dynamics in a self-starting passively mode-locked fiber-based soliton laser, *Appl. Phys. Lett.* **63**, 2857-2859 (1993).

5. J. M. Dudley, C. M. Loh, J. D. Harvey, Stable and unstable operation of a mode-locked argon laser, *Quantum Semiclass. Opt.* **8** 1029-1039 (1996).

6. C. Hönninger, R. Paschotta, F. Morier-Genoud, M. Moser, and U. Keller, Q-switching stability limits of continuous wave passive mode locking, *J. Opt. Soc. Am. B* **16**, 46-56 (1999).

7. N. Akhmediev, A. Ankiewicz (Eds.), *Dissipative Solitons.* Lecture Notes in Physics, Springer-Verlag, 2005.

8. P. Grelu, N. Akhmediev, Dissipative solitons for mode-locked lasers, *Nat. Photon.* **6**, 84–92 (2012).

9. P. Grelu (Ed.) *Nonlinear Optical Cavity Dynamics: From Microresonators to Fiber Lasers.* Wiley, 2016.

10. S.K. Turitsyn, N.N. Rozanov, I.A. Yarutkina, A.E. Bednyakova, S.V. Fedorov, O.V. Shtyrina, M.P. Fedoruk. Dissipative solitons in fiber lasers, *Phys. Usp.* **59**, 642-668 (2016).

11. M. B. Flynn, L. O'Faolain, T. F. Krauss, An experimental and numerical study of Q-switched mode-locking in monolithic semiconductor diode lasers, *IEEE J. Quant. Electron.* **40**, 1008-1013 (2004).

12. A. Schlatter, S. C. Zeller, R. Grange, R. Paschotta, U. Keller, Pulse-energy dynamics of passively mode-locked solid-state lasers above the Q-switching threshold, *J. Opt. Soc. Am. B* **21**, 1469-1478 (2004).

13. C. Lecaplain, P. Grelu, J. M. Soto-Crespo, N. Akhmediev, Dissipative rogue waves generated by chaotic pulse bunching in a mode-locked laser, *Phys. Rev. Lett.* **108**, 233901 (2012).
15

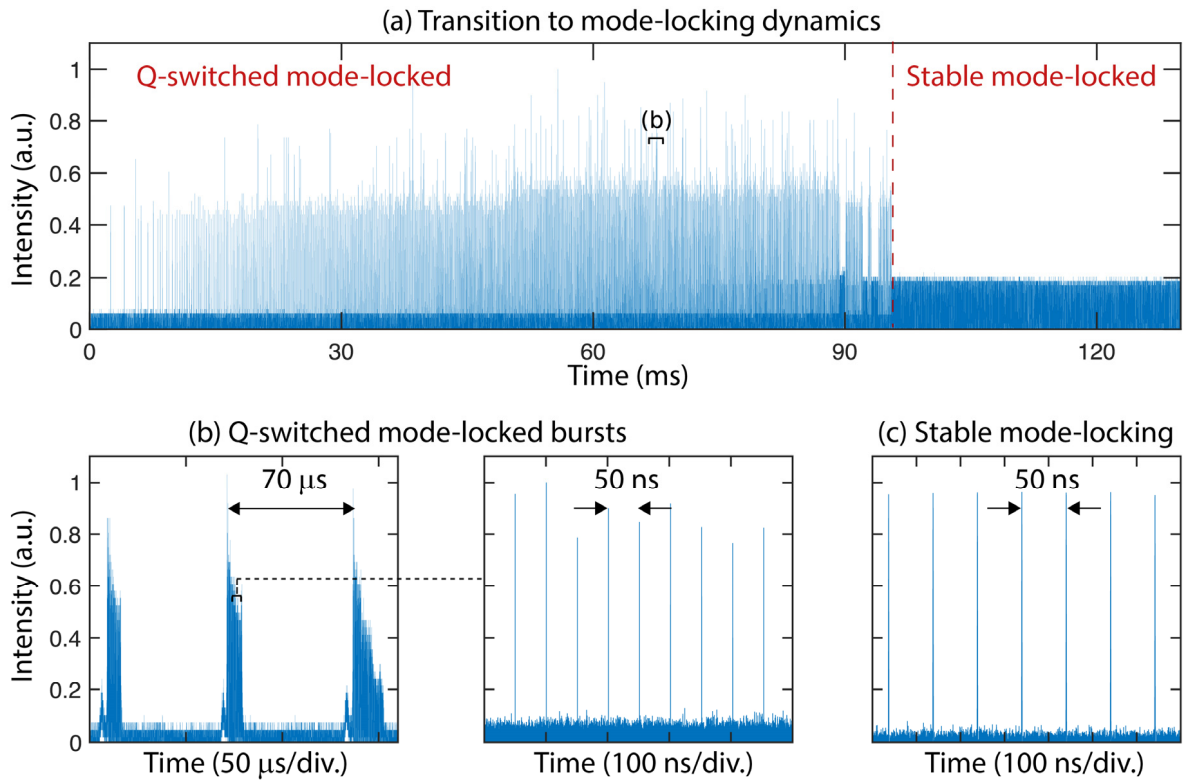

**Figure 1**. Direct photodetector measurement of transient laser dynamics. (a) Results recorded with reduced bandwidth detection of 20 MHz illustrating the ~100 ms transient regime of Q-switched mode-locked operation before stable mode-locking. The total time record shown corresponds to 2.6 million round trips. (b) Results from separate measurements using 30 GHz bandwidth detection, showing how the laser output during the Q-switched mode-locked regime consists of transient bursts of temporal width ~10 µs separated by ~70 µs period. The expanded view shows how unstable mode-locked pulses at 50 ns period are generated under each burst. (c) Results using 30 GHz detection but in the stable mode-locked regime, showing a regular train of pulses with constant amplitude.

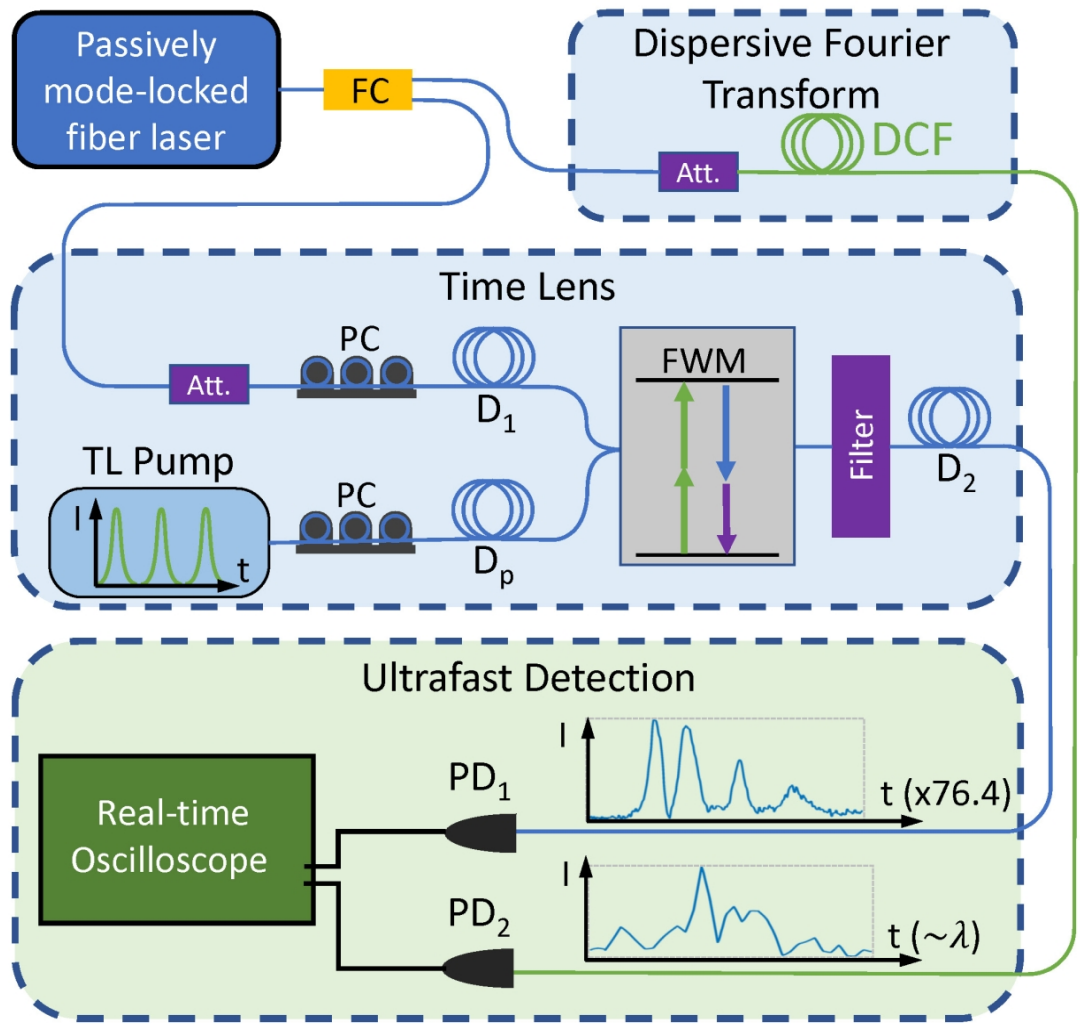

**Figure 2**. Setup used for real-time characterization. Pulses from a 20 MHz passively mode-locked fiber laser are split into two paths in a fiber coupler (FC) and sent to a DFT real time spectral setup and a time-lens. The DFT used 850 m of dispersion compensating fiber (DCF) to perform time to wavelength conversion, with detection using a 35 GHz bandwidth photodiode ($PD_2$). An optical spectrum analyzer was also used for control measurements of the spectrum. The time lens uses two dispersive propagation steps ($D_1$ and $D_2$), one on each side of a silicon waveguide stage that applies a quadratic temporal phase through four wave mixing (FWM) with linearly-chirped pump pulses generated from a 100 MHz femtosecond pulse fiber laser (TL Pump) after stretching in a dispersive stretching fiber $D_p$. The time lens signal is extracted from the FWM spectrum by an optical filter and was measured using a 13 GHz bandwidth photodiode ($PD_1$). Both $PD_1$ and $PD_2$ were input to a 30 GHz channel of the real-time oscilloscope. Values for dispersion modules $D_1$, $D_2$ and $D_p$ in the time lens are given in Materials and Methods. The inputs to both the DFT and time lens were attenuated (Att.) to avoid nonlinear effects and polarization control (PC) was needed to ensure optimal signal from the time-lens.

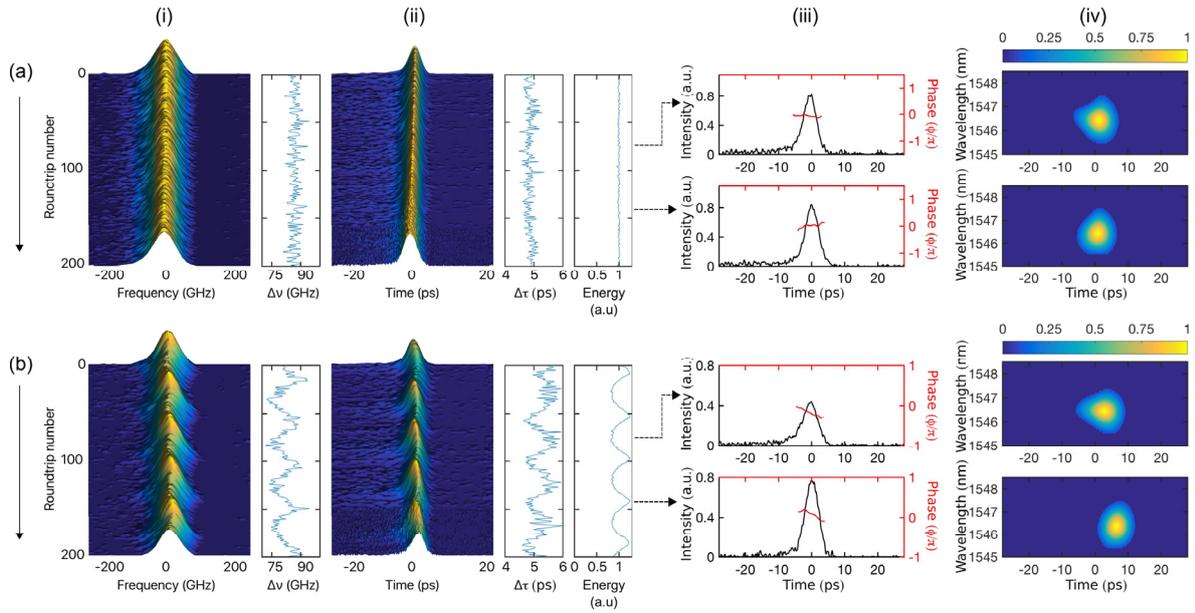

**Figure 3**. Results over 200 round trips showing real-time spectral and temporal characterization for (a) stable and (b) breathing mode-locking regimes. (i) Measured spectral intensity with the extracted spectral width shown in the right subfigure. (ii) Measured temporal intensity with the extracted temporal width and energy shown in the right subfigures. (iii) Corresponding temporal intensity and phase extracted using the Gerchberg-Saxton algorithm for pulses at points indicated by arrows. For each of the extracted pulses in (iii), the plots in (iv) show the calculated wavelength-time spectrogram.

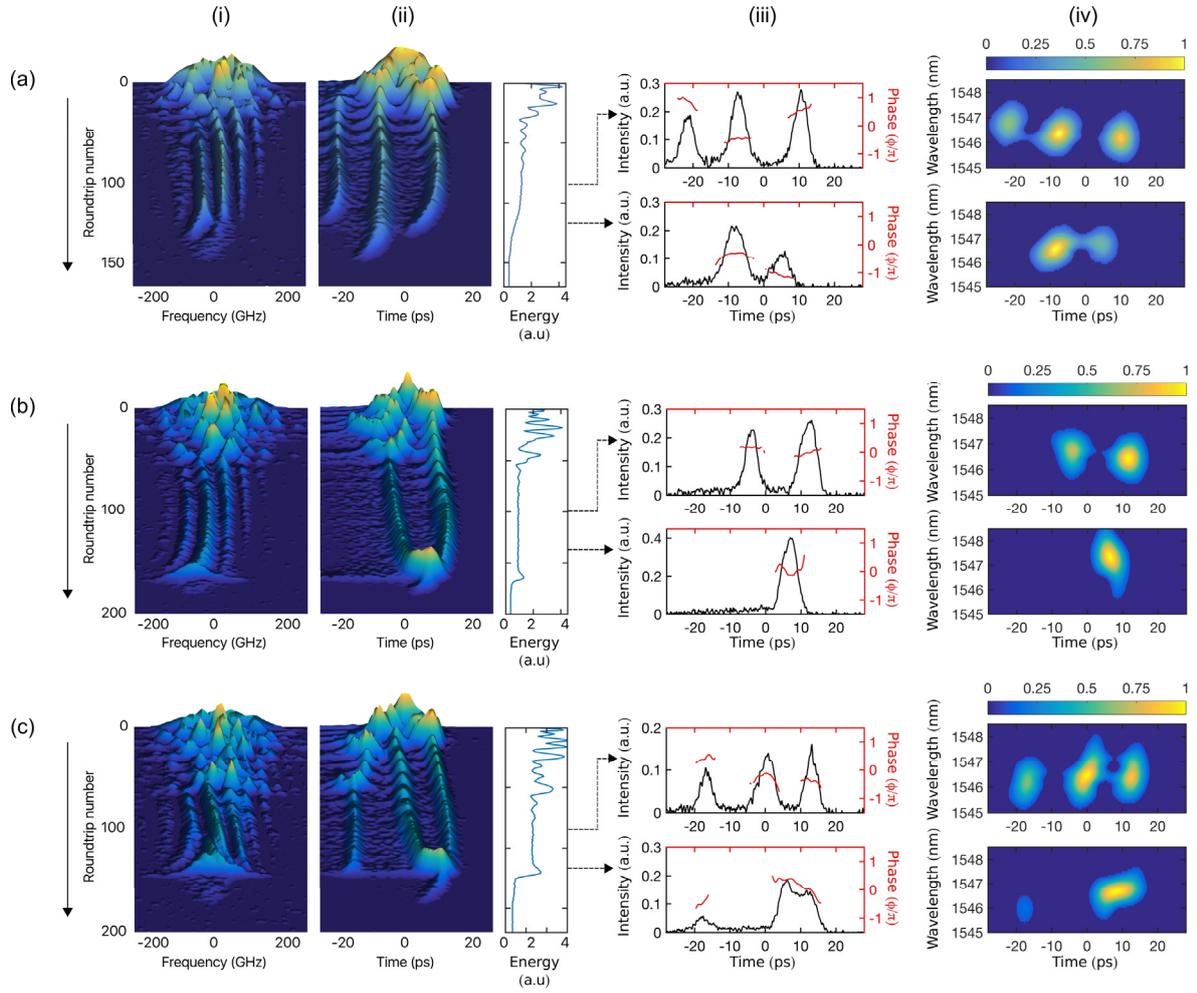

**Figure 4**. Results over 200 round trips showing real-time spectral and temporal characterization for (a) a non-interacting triplet of three solitons, (b) more complex break-up and collision dynamics of a soliton double pulse and (c) a combination of a single non-interacting soliton and a two pulse collision. (i) Measured spectral intensity. (ii) Measured temporal intensity. The right subfigure shows the integrated energy. (iii) Corresponding temporal intensity and phase extracted using the Gerchberg-Saxton algorithm for pulses at points indicated by arrows. For each of the extracted pulses in (iii), the plots in (iv) show the calculated wavelength-time spectrogram.

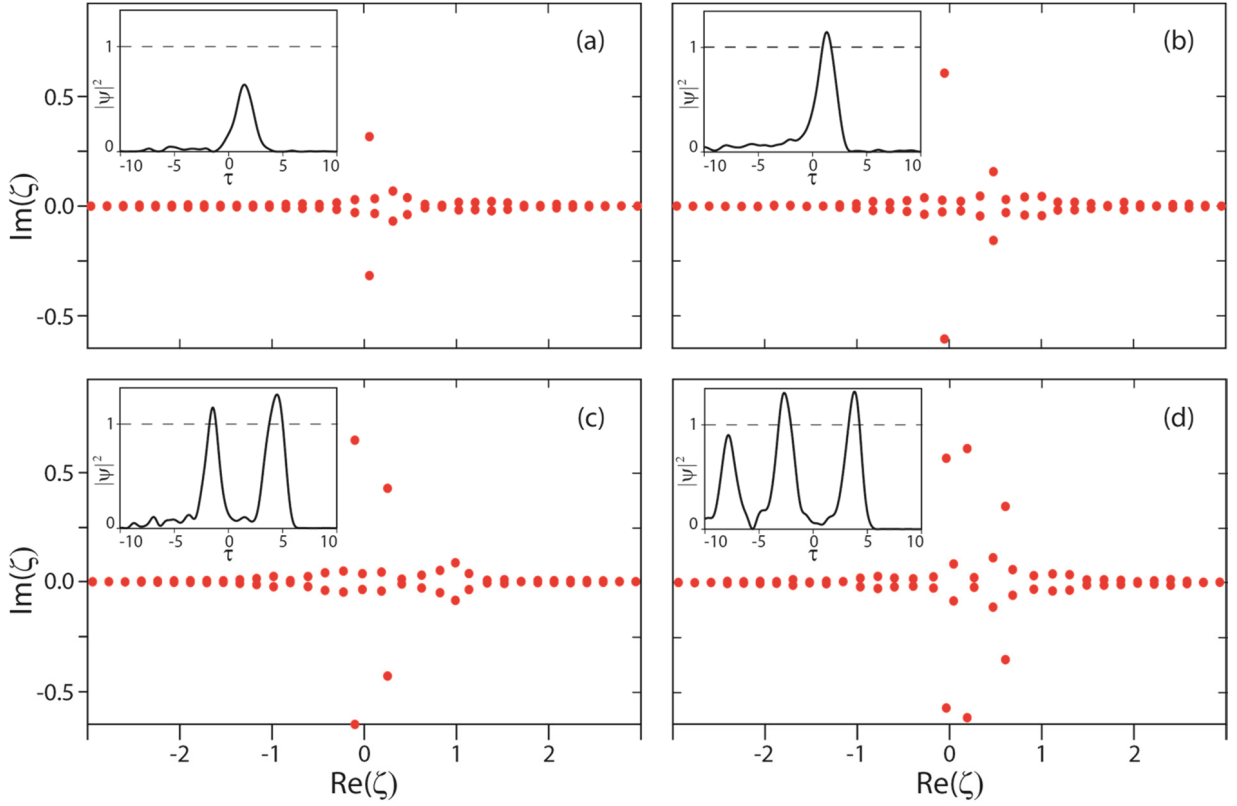

**Figure 5**. Results applying the scattering transform to intensity and phase measurements of a selection of measured dissipative solitons. In each case the red dots show the calculated discrete eigenspectrum corresponding to the normalized pulse intensity profiles shown as insets. The results in Fig 5(a) and (b) correspond to the two points in the dissipative soliton breather cycle shown in Fig. 3(b). As the pulse intensity varies below and above the unity value of a normalized fundamental soliton, the retrieved eigenvalues Im($\zeta$) varies below and above ±0.5$i$. (c) The double soliton case in Fig. 4(b) where we see two discrete eigenvalues with Im($\zeta$) ~ ±0.5$i$. (d) The three soliton case in Fig. 4(a) where we see three discrete eigenvalues with Im($\zeta$) around ±0.5$i$. (Note that the intensity profiles plotted correspond to results shown in Fig. 3 and 4 but have been smoothed for clarity when shown as insets.) The dashed line in the insets shows the intensity of an ideal soliton.